\documentclass[aps,prl,superscriptaddress,showpacs,showkeys,manuscript]{revtex4}
\usepackage{CJK}%
\usepackage{color}%
\usepackage{amsmath,bm}%
\usepackage{graphicx}%

\begin{document}
\begin{CJK*}{GBK}{}
\title{Reconstructed primary fragments and symmetry energy, temperature and density of the fragmenting source in $^{64}$Zn + $^{112}$Sn at 40 MeV/nucleon}
\author{X. Liu}
\email[E-mail at:]{liuxingquan@impcas.ac.cn}
\affiliation{Institute of Modern Physics, Chinese Academy of Sciences, Lanzhou, 730000, China}
\affiliation{University of Chinese Academy of Sciences, Beijing 100049, China}
\author{W. Lin}
\affiliation{Institute of Modern Physics, Chinese Academy of Sciences, Lanzhou, 730000, China}
\affiliation{University of Chinese Academy of Sciences, Beijing 100049, China}
\author{R. Wada}
\email[E-mail at:]{wada@comp.tamu.edu}
\affiliation{Institute of Modern Physics, Chinese Academy of Sciences, Lanzhou, 730000, China}
\author{M. Huang}
\affiliation{Institute of Modern Physics, Chinese Academy of Sciences, Lanzhou, 730000, China}
\author{S. Zhang}
\affiliation{Institute of Modern Physics, Chinese Academy of Sciences, Lanzhou, 730000, China}
\affiliation{University of Chinese Academy of Sciences, Beijing 100049, China}
\author{P. Ren}
\affiliation{Institute of Modern Physics, Chinese Academy of Sciences, Lanzhou, 730000, China}
\affiliation{University of Chinese Academy of Sciences, Beijing 100049, China}
\author{Z. Chen}
\affiliation{Institute of Modern Physics, Chinese Academy of Sciences, Lanzhou, 730000, China}
\author{J. Wang}
\affiliation{Institute of Modern Physics, Chinese Academy of Sciences, Lanzhou, 730000, China}
\author{G. Q. Xiao}
\affiliation{Institute of Modern Physics, Chinese Academy of Sciences, Lanzhou, 730000, China}
\author{R. Han}
\affiliation{Institute of Modern Physics, Chinese Academy of Sciences, Lanzhou, 730000, China}
\author{J. Liu}
\affiliation{Institute of Modern Physics, Chinese Academy of Sciences, Lanzhou, 730000, China}
\author{F. Shi}
\affiliation{Institute of Modern Physics, Chinese Academy of Sciences, Lanzhou, 730000, China}
\author{M. R. D. Rodrigues}
\affiliation{Instituto de F\'{\i}sica, Universidade de S\~{a}o Paulo, Caixa Postal 66318, CEP 05389-970, S\~{a}o Paulo, SP, Brazil}
\author{S. Kowalski}
\affiliation{Institute of Physics, Silesia University, Katowice, Poland.}
\author{T. Keutgen}
\affiliation{FNRS and IPN, Universit\'e Catholique de Louvain, B-1348 Louvain-Neuve, Belgium}
\author{K. Hagel}
\affiliation{Cyclotron Institute, Texas A$\&$M University, College Station, Texas 77843}
\author{M. Barbui}
\affiliation{Cyclotron Institute, Texas A$\&$M University, College Station, Texas 77843}
\author{A. Bonasera}
\affiliation{Cyclotron Institute, Texas A$\&$M University, College Station, Texas 77843}
\affiliation{Laboratori Nazionali del Sud, INFN,via Santa Sofia, 62, 95123 Catania, Italy}
\author{J. B. Natowitz}
\affiliation{Cyclotron Institute, Texas A$\&$M University, College Station, Texas 77843}
\author{H. Zheng}
\affiliation{Cyclotron Institute, Texas A$\&$M University, College Station, Texas 77843}
\affiliation{Physics Department, Texas A$\&$M University, College Station, Texas 77843}
\date{\today}

\begin{abstract}
Symmetry energy, temperature and density at the time of the intermediate mass fragment formation are determined in a self-consistent manner, using the experimentally reconstructed primary hot isotope yields and anti-symmetrized molecular dynamics (AMD) simulations. The yields of primary hot fragments are experimentally reconstructed for multifragmentation events in the reaction system $^{64}$Zn + $^{112}$Sn at 40 MeV/nucleon. Using the reconstructed hot isotope yields and an improved method, based on the modified Fisher model, symmetry energy values relative to the apparent temperature, $a_{sym}/T$, are extracted. The extracted values are compared with those of the AMD simulations, extracted in the same way as those for the experiment, with the Gogny interaction with three different density-dependent symmetry energy terms. The $a_{sym}/T$ values change according to the density-dependent symmetry energy terms used. Using this relation, the density of the fragmenting system is extracted first. Then symmetry energy and apparent temperature are determined in a self consistent manner in the AMD model simulations. Comparing the calculated $a_{sym}/T$ values and those of the experimental values from the reconstructed yields, $\rho /\rho_{0} = 0.65 \pm 0.02 $, $a_{sym} = 23.1 \pm 0.6$ MeV and $T= 5.0 \pm 0.4$ MeV are evaluated for the fragmenting system experimentally observed in the reaction studied.

\end{abstract}


\maketitle
\end{CJK*}

\section*{I. Introduction}

Nuclear symmetry energy, a part of the equation of state (EoS) in the nuclear matter equation, has been extensively studied in the last three decades. The symmetry energy relates to many subjects such as in nuclear astrophysics, nuclear structure, and nuclear reactions. Its property determination is a key objective in laboratory experiments~\cite{Lattimer04,BALi08}. Investigations of the symmetry energy, especially focusing on its density dependence, have been conducted using many observables such as isotopic yield ratios~\cite{Tsang2001}, isospin diffusion~\cite{Tsang2004}, neutron-proton emission ratios~\cite{Famiano2006}, giant monopole resonances~\cite{Li2007}, pygmy dipole resonances~\cite{Klimkiewicz2007}, giant dipole resonances~\cite{Trippa2008}, collective flows~\cite{zak2012} and isoscaling~\cite{Xu2000,Tsang2001_1,Huang2010}. Different observables may probe the properties of the symmetry energy at different densities and temperatures.

In a theoretical work of the EoS study, Wiringa et al.~\cite{Wiringa88} pointed out that the density dependence of the symmetry energy may have different slope parameters in different higher density regions. When a three body interaction is taken into account, the symmetry energy shows a significant softening at $\rho/\rho_0 \sim 2-3$, hardening again at $\rho/\rho_0 \sim 5$ and then shows an asymptotic soft trend for the higher density. Therefore it is important to know not only the values of the symmetry energy and slope parameter or the exponent of the density dependent terms, but also the density and temperature of the system when the values are evaluated.

In one of our previous works, the density dependence of the symmetry energy at low densities were experimentally studied in several heavy ion reactions at 47 MeV/nucleon, using the light particles ($Z=1,2$) from the intermediate velocity source as the probe~\cite{Wada2012}. In that study the temperature in the region 5$-$10 MeV was evaluated from the double ratio thermometer and the density of $0.03\leq \rho/\rho_{0}\leq 0.2$ was extracted from the coalescence technique. In the sampled density and temperature intervals, symmetry energies were derived and nonzero symmetry energies were obtained at low densities.
However in the quasiparticle approaches, such as Skyrme Hartree-Fock and relativistic mean field models or Dirac-Brueckner Hartree-Fock calculations, the symmetry energy tends to zero at low densities~\cite{BALi08,DiToro10,Sammarruca10}. This significant experimentally observed symmetry energy deviation at low densities from those of the quasiparticle predictions can be attributed to the cluster formation which dominates the structure of low-density symmetric matter at low temperatures, in accordance with the mass action law.

In violent heavy ion collisions at intermediate energy regime (20 $\leq E_{inc} \leq$ a few hundred MeV/nucleon), intermediate mass fragments (IMFs) are copiously produced through multifragmentation processes. Nuclear multifragmentation, which in general, can be divided into stages, i.e., the dynamical compression and expansion of the fragmenting source, and the formation of primary hot fragments, was predicted a long time ago~\cite{Bohr36} and has been studied extensively following the advent of 4$\pi$ detectors~\cite{Borderie08,Gulminelli06,Chomaz04}. Nuclear multifragmentation occurs when a large amount of energy is deposited in a finite nucleus, and thus it provides important information on the properties of the hot nuclear matter equation of state.


To model the multifragmentation process, a number of different models have been developed in two distinct scenarios. One is based on a transport model, in which nucleon propagation in a mean field and nucleon-nucleon collisions under Pauli-blocking are two main physical ingredients.  Various transport models have been coded, since Boltzmann-Uehling-Uhlenbeck (BUU) model~\cite{Aichelin85} was first proposed in 1980s, which is a test particle based Monte Carlo transport model. Vlasov-Uehling-Uhlenbeck model (VUU)~\cite{Kruse85}, Boltzmann-Nordheim-Vlasov model (BNV)~\cite{Baran02} are formulated slightly differently with the same concept.  Stochastic mean field (SMF) model~\cite{Colonna98,Baran12,Gagnon12} is also a test particle based model, but with fluctuations in multifragmentation process. Instead of using the test particles, Gaussian wave packets are introduced in describing the nucleons such as quantum molecular dynamics model (QMD)~\cite{Peilert89,Aichelin91,Lukasik93}. Constrained molecular dynamics(CoMD) model~\cite{Papa01,Papa05,Papa07,Papa09} and improved quantum molecular dynamics model (ImQMD)~\cite{Wang02,Wang04,Zhang05,Zhang06,Zhang12} are based on QMD, but an improved treatment is made on the Pauli blocking during the time evolution of the reaction. Fermionic molecular dynamics(FMD)~\cite{Feldmeier90} and anti-symmetrized molecular dynamics (AMD)~\cite{Ono96,Ono99,Ono02} are most sophisticated models, in which the Pauli principle is taken into account in an exact manner in the time evolution of the wave packet and nucleon-nucleon collisions.
Most of them can account reasonably well for many characteristic properties experimentally observed. On the other hand statistical multifragmentation models such as microcanonical Metropolitan Monte Carlo model (MMMC)~\cite{Gross90,DAgostino99} and statistical multifragmentation model(SMM)~\cite{DAgostino99,Bondorf85,Bondorf95,Botvina95,DAgostino96,Scharenberg01,Bellaize02,Avdeyev02,Ogul12},  based on a quite different assumption from the transport models, can also describe many experimental observables well. The statistical models use a freeze-out concept. The multifragmentation is assumed to take place in equilibrated nuclear matter described by parameters, such as size, neutron/proton ratio, density and temperature. In recent analyses the parameters are optimized to reproduce the experimental observables of the final state. In contrast, the transport models do not assume any chemical or thermal equilibration. Nucleons travel in a mean field experiencing nucleon-nucleon collisions subject to the Pauli principle. Fragmentation mechanisms are determined by the evolutions of the wave pockets or nucleons in the phase space, which also differ from those of the statistical models.

One of the complications one has to face when comparing the experimental observables to the model predictions in either dynamical or statistical models, is the secondary decay process. When fragments are formed in a multifragmentation process, many of them can be in excited states and cool down by evaporation processes before they are detected experimentally~\cite{Marie98,Hudan03,Rodrigues13,Lin14_1,Lin14_2}.
Here the fragments at the time of formation are called "primary" fragments. Those observed after the cooling process are called the "secondary" or "final" fragments.
Multifragmentation process is a very fast process which occurs in an order of 50-100 fm/c in the intermediate energy heavy ion collisions, whereas the secondary decay process is much slower. Therefore the secondary cooling process may significantly alter the fragment yield distributions of the primary isotopes~\cite{Huang10_1,Huang10_2,Chen10}. Even though the statistical decay process itself is rather well understood and well coded, it is not a trivial task to combine it with a dynamical code. That is because the statistical evaporation codes assume the nuclei at thermal equilibrium with normal nuclear density and shapes. However these conditions are not guaranteed for fragments when they are formed in the multifragmentation process.

In order to avoid this complication and make the comparisons between results from the experimental data and different models more straight forward, we proposed a method in which the primary hot fragment yields are reconstructed experimentally. The method utilizes a kinematic focusing of the evaporated particles along the precursors of IMFs.
In Fermi energy heavy ion collisions, light particles are emitted at different stages of the reaction and from different sources during the evolution of the collisions. Those from an excited isotope are kinematically focused into a cone centered along the isotope direction. The kinematical focusing technique uses this nature.
Details of the experiment, the kinematical focusing technique and the results are presented in Refs.~\cite{Rodrigues13,Lin14_1}.

\begin{figure*}[htb]
\includegraphics[scale=0.7]{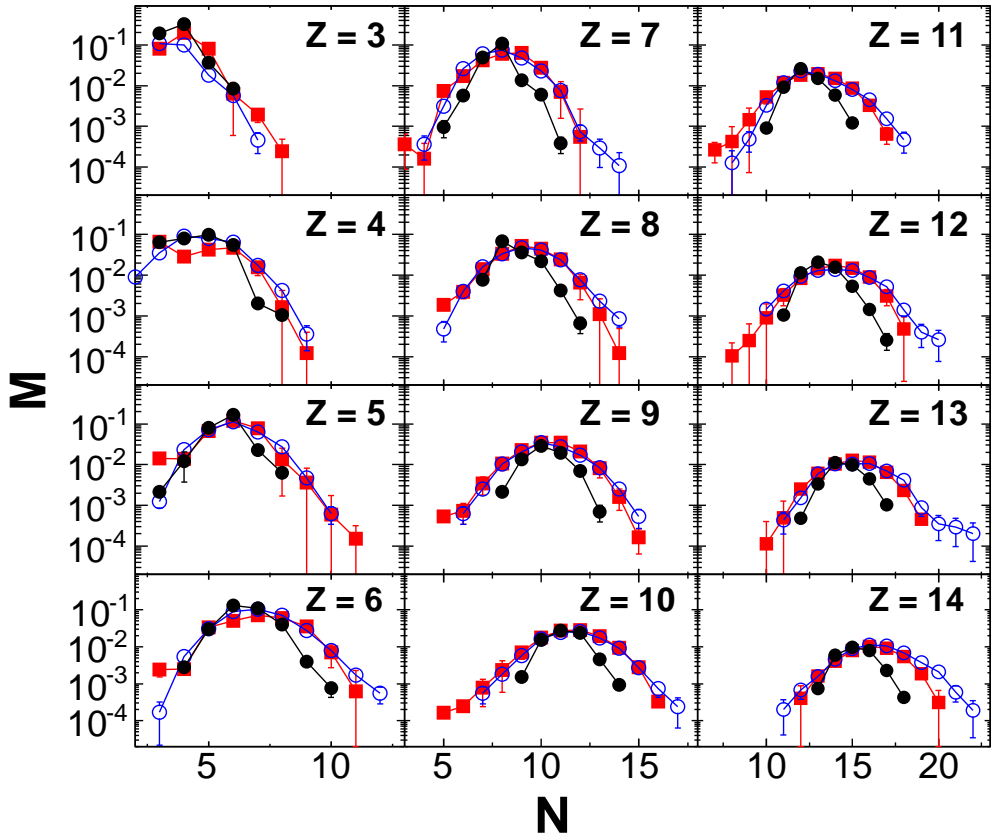}
\caption{\footnotesize
(Color online) Isotopic multiplicity distributions of experimental cold fragments (dots), reconstructed hot fragments (closed squares) as well as AMD primary hot fragments (circles) as a function of fragments mass number $A$ for a given charge $Z$, which is indicated in the figure. In the AMD simulations, g0AS interaction is used.
}			
\label{fig:Isotopic_distribution}
\end{figure*}

In that work, the events triggered by IMFs in the experiment are "inclusive", but they belong to a certain class of events. In order to determine the event class taken in the experiment, AMD simulations are used to evaluate the impact parameter range sampled.
Firstly the impact parameter distributions, corresponding violent, semi-violent, semi-peripheral and peripheral collisions are calculated.  The violence of the reaction for each event in the AMD simulation is determined in the same way as our previous work~\cite{Wada04}. Then the impact parameter distribution of the events triggered by the IMFs at $20^{\circ}$ is calculated and compared to those corresponding to the different violence. The distribution is very similar to those of the semi-violent collisions, in which the majority of the events originates from the impact parameter range of $0-8$ fm.
Therefore in the following analyses, the comparisons of the extracted parameters from the experimentally reconstructed isotope yields are made with those of the AMD simulations in the impact parameter range of $0-8$ fm.
In Fig.~\ref{fig:Isotopic_distribution}. the results of the multiplicity distributions of the experimental cold and reconstructed hot isotopes are shown, together with those of the primary isotopes simulated by the AMD calculations. The reconstructed isotope multiplicities are reasonably well reproduced by the primary isotope distribution of the AMD simulation. In Refs.~\cite{Lin14_1,Lin14_2}, we studied the properties of the fragmenting system through the symmetry energy coefficient relative to the temperature, $a_{sym}/T$.
In the study the $a_{sym}/T$ values were extracted in a simpler formalism, utilizing three isobars of the reconstructed primary hot fragments with $I=N-Z=-1,1$ and 3. This article presents an improved method to calculate the $a_{sym}/T$ values, in which the mass dependence of the temperature is taken into account as an apparent temperature.
This method has been applied recently to the simulated AMD events of the very central collisions for $^{40}Ca+^{40}Ca$ at 35 MeV/nucleon~\cite{Liu14}. A self-consistent determination of density, symmetry energy and temperature described in Refs.~\cite{Lin14_1,Lin14_2} was also employed there.
In this work the same procedure following Ref.~\cite{Liu14} is applied to the experimentally reconstructed isotope yields of $^{64}Zn + ^{112}Sn$ at 40 MeV/nucleon to study the characteristic properties of the hot nuclear matter in the multifragmenting system.

This article is organized as follows.
In Sec.II we describe the improved method to determine the symmetry energy coefficient relative to the temperature, $a_{sym}/T$, utilizing all isotope yields. In Sec.III, a self-consistent determination of density, symmetry energy and temperature is discussed. In Sec.IV, the mass dependent apparent temperature is studied. Finally, a summary is given in Sec.V.

\section*{II. Extraction of $a_{sym}/T_{0}$ values}

In order to make a connection between the symmetry energy in a model and the experimentally reconstructed primary hot isotope yields in Fig.~\ref{fig:Isotopic_distribution}, the Modified Fisher Model (MFM) is employed~\cite{Fisher1967,Minich1982,Hirsch1984,Bonasera2008}.
MFM has been used to study the characteristic properties of the hot nuclear matter in the previous works~\cite{Bonasera2008,Huang10_1,Huang10_2,Huang10_3,Lin14_1,Lin14_2,Liu14}.
In the framework of MFM, the yield of an isotope with $I=N-Z$ and $A$ ($N$ neutrons and $Z$ protons) produced in a multifragmentation reaction, can be given as
\begin{equation}
\begin{split}
Y(I,A) =& Y_{0} \cdot A^{-\tau}exp[\frac{W(I,A)+\mu_{n}N+\mu_{p}Z}{T}\\
&+S_{mix}(I,A)].
\end{split}
\label{eq:eq_MFM}
\end{equation}
Using the generalized Weizs$\ddot{a}$cker-Bethe semiclassical mass formula~\cite{Weizsacker1935,Bethe1936}, $W(I,A)$ can be approximated as
\begin{equation}
\begin{split}
W(I,A) =& a_{v}A- a_{s}A^{2/3}- a_{c}\frac{Z(Z-1)}{A^{1/3}}\\
&-a_{sym}\frac{I^{2}}{A}
- a_{p}\frac{\delta}{A^{1/2}},\\
\delta =& - \frac{(-1)^{Z}+(-1)^{N}}{2}.
\end{split}
\label{eq:eq_WB}
\end{equation}
In Eq.\eqref{eq:eq_MFM}, $A^{-\tau}$ and $S_{mix}(I,A)=Nln(N/A)+Zln(Z/A)$ originate from the increases of the entropy and the mixing entropy at the time of the fragment formation, respectively. $\mu_{n}$ ($\mu_{p}$) is the neutron (proton) chemical potential. $\tau$ is the critical exponent. In this work, the value of $\tau=2.3$ is adopted from the previous studies~\cite{Bonasera2008}. Since we apply this formulation for the primary hot fragments, the coefficients, $a_{v}$, $a_{s}$, $a_{sym}$, $a_{p}$ and the chemical potentials, are generally temperature and density dependent, even though these dependencies are not shown explicitly.

In this formulation a constant volume process at an equilibrium is assumed in the free energy, and therefore the term "symmetry energy" is used throughout this work, following Ref.~\cite{Marini2012}. If one assumes a constant pressure at the equilibrium process~\cite{Sobotka2011}, the therm "symmetry enthalpy" should be used. Experimentally, whether the equilibrium process takes place at constant pressure or volume can not be determined, and thus we use "symmetry energy" through out the paper, keeping in mind the ambiguity~\cite{Marini2012}.

In the previous analyses~\cite{Lin14_1,Lin14_2,Huang10_1,Huang10_2,Chen10}, the temperature in Eq.\eqref{eq:eq_MFM} was assumed to be identical to the temperature of the fragmenting source and treated as a constant for all isotopes. However as seen in Ref.~\cite{Liu14}, this temperature turns out to be fragment mass dependent. This mass dependence on the temperature was not recognized in these previous analyses, just because the mass dependence was masked by the larger error bars. However in this improved method, the error bars become small and the mass dependence becomes evident. In order to take into account this mass dependence of the temperature in Eq.\eqref{eq:eq_MFM}, the temperature $T$ is replaced by an apparent temperature
$T(A)=T_0(1-kA)$. $T_{0}$ is the temperature of the fragmenting source and k is a constant. As discussed in Ref.~\cite{Liu14}, this mass dependence of the apparent temperature is attributed to the system size effect.

In order to study the density, temperature and symmetry energy in the fragmenting source, the improved MFM of Eq.\eqref{eq:eq_MFM} is utilized to calculate the $a_{sym}/T_{0}$ value, which is extracted from the available isotope yields. Since the $a_{sym}/T_{0}$ value in Eqs.\eqref{eq:eq_MFM} and \eqref{eq:eq_WB}  depends on 5 parameters, $a_{v}$, $a_{s}$, $a_{c}$, $a_{p}$ and $\Delta \mu$ ($\Delta \mu=\mu_n-\mu_p$), the optimization process of these parameters is divided into the following three steps to minimize the ambiguity of each parameter. For a given $k$ value,

\begin{enumerate}
\item{Optimize $\Delta\mu/T_{0}$ and $a_c/T_{0}$ values from mirror isobars.}
\item{Optimize $a_v/T_{0}$, $a_s/T_{0}$ and $a_p/T_{0}$ values from $N=Z$ isotopes.}
\item{Using extracted parameters in step (1) and step (2), $a_{sym}/T_{0}$ values are extracted from all available isotopes. Comparing the extracted $a_{sym}/T_{0}$ values from the AMD simulations with three different interactions, the density of the fragmenting source is extracted. Using this density, the value of the symmetry energy coefficient, $a_{sym}$, for each interaction is determined. The temperature is then extracted following the relation, $T_{0}=a_{sym}/(a_{sym}/T_{0})$.
    }
\end{enumerate}

It is expectable that if the $k$ value is properly selected which means the mass dependence is well considered, a constant $T_{0}$ is obtained. Since the $k$ value is small as seen below, we perform the optimization of the parameter $k$ in an iterative manner, that is, in the first round $k=k_{1}=0$ is set in $T(A)=T_0(1-kA)$ and calculate the temperature as a function of $A$, using steps (1)-(3). From this plot a new $k_{1}'$ value is extracted from the slope. In the second round, $k=k_{2}=k_{1}+\frac{1}{2}k_{1}'$ is used for the steps (1)-(3) and a new $k_{2}'$ value is extracted. If the new $k_{2}'$ value is 0 within a given error range, the iteration stops and the $k_{2}$ value is fixed as the mass dependent parameter of the apparent temperature and $T_{0}$ value is determined. Otherwise the iteration continues.

These procedures are applied individually for the reconstructed isotope yields and the AMD simulated events with interactions having different density dependencies of the symmetry energy term, i.e., the standard Gogny interaction which has an asymptotic soft symmetry energy (g0), the Gogny interaction with an asymptotic stiff symmetry energy (g0AS) and the Gogny interaction with an asymptotic super-stiff symmetry energy (g0ASS)~\cite{Ono99,Ono03}. To keep consistent with experimental isotope selections, for AMD primary hot fragments, an approximate window is employed, in which the multiplicity of the IV source component is calculated by integrating the energy spectra over $E > 5$ MeV/nucleon and between $5^{\circ}<\theta <25^{\circ}$ in the laboratory frame in order to minimize the contribution from the projectile-like and the target-like sources, based on the moving source analysis~\cite{Lin14_1,Lin14_2}.

\begin{figure}
\includegraphics[scale=0.45]{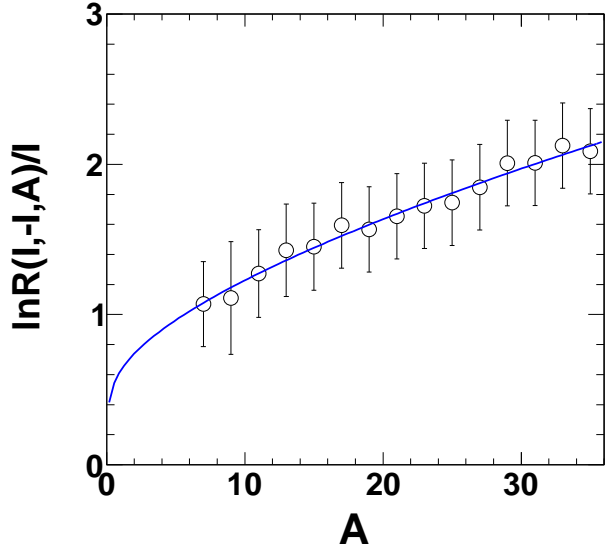}
\caption{\footnotesize (color online)
$ln[R(I,-I,A)]/I$ versus $A$ for mirror nuclei with $I=1$ for the case of the reconstructed isotope yields. The curve is the fit result of Eq.\eqref{eq:Coulomb} for $k=0$. The extracted values of $\Delta\mu/T_{0}$ and $a_c/T_{0}$ are given in the third and fifth columns of Table~\ref{table:parameters}.
}
\label{fig:fig1}
\end{figure}

Details of each step are described below.
In the step (1), following Ref.~\cite{Huang10_1}, the isotope yield ratio between isobars with $I+2$ and $I$, $R(I+2,I,A)$, is utilized, which is
\begin{eqnarray}
R(I+2,I,A) = Y(I+2,A)/Y(I,A)    \nonumber\\
=  exp\{[\mu_{n}- \mu_{p}+ 2a_{c}(Z-1)/A^{1/3}-\nonumber\\ 4a_{sym}(I+1)/A-\delta(N+1,Z-1) \nonumber\\
- \delta(N,Z)]/[T_{0}(1-kA)] + \Delta(I+2,I,A)\},
\label{eq:eq_RI2}
\end{eqnarray}
where $Y(I,A)$ is the yield of isotopes with $I$ and $A$, and $\Delta(I+2,I,A)=S_{mix}(I+2,A) - S_{mix}(I,A)$.
When the above equation is applied for a pair of mirror nuclei of odd mass isotopes with $I = -I$ and $I$, the symmetry energy term, pairing term and mixing entropy terms drop out and the following equation is obtained.
\begin{equation}
\ln[R(I, -I, A)]/I=[\Delta\mu+a_{c}(A-1)/A^{1/3}]/[T_{0}(1-kA)].
\label{eq:Coulomb}
\end{equation}
For available mirror isobars with I=1 and -1, $\Delta\mu/T_{0}$ and $a_c/T_{0}$ are optimized in Eq.\eqref{eq:Coulomb}. The $\ln[R(I, -I, A)]/I$ values and the fit result for $k=0$ is shown in Fig.\ref{fig:fig1} for the case of the reconstructed isotope yields. Similar results are obtained for the AMD simulated events.

\begin{figure}
\includegraphics[scale=0.45]{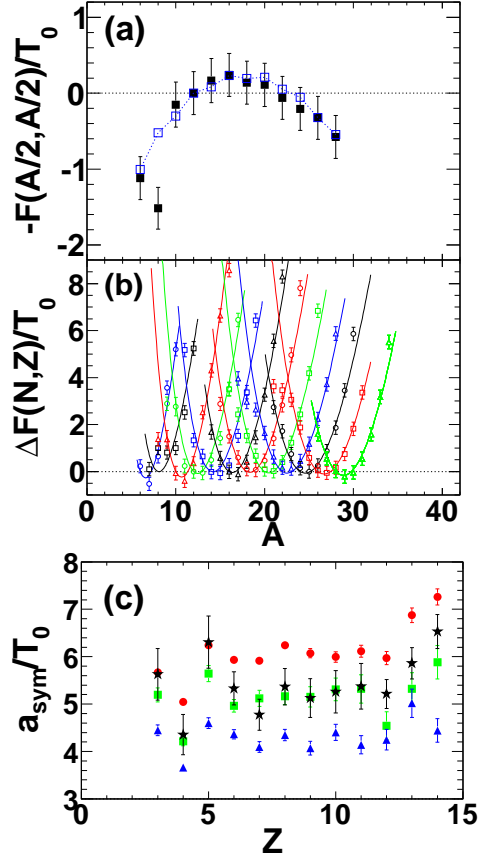}
\caption{\footnotesize (color online)
(a) Calculated ratio of free energy relative to $T_0$ for $N=Z$ isotopes from the reconstructed isotope yields (solid squares). Open squares represent the fit using Eq.\eqref{eq:eq5}. The parameters extracted are given in Table.~\ref{table:parameters}. (b) Calculated $\frac{\Delta F(N,Z)}{T}$ values (symbols) and quadratic fits (curves) using Eq.\eqref{eq:eq7} for $Z=3$ to 14 for the reconstructed isotope yields. The same symbols are used for isotopes with a given $Z$. (c) Extracted $a_{sym}/T_{0}$ values from (b) for the reconstructed (stars), g0 (dots), g0AS (squares) and g0ASS (triangles). All values are evaluated at the first round $k=0)$.
}
\label{fig:fig2}
\end{figure}

In the step (2) we apply Eq.\eqref{eq:eq_MFM} to the $N=Z$ isotopes with the extracted $\Delta\mu/T_{0}$ and $a_c/T_{0}$ values in the step (1).
For $N = Z = A/2$ isotopes, the free energy relative to the temperature can be calculated from Eq.\eqref{eq:eq_MFM} and Eq.\eqref{eq:eq_WB} without the symmetry energy term as
\begin{equation}
\begin{split}
-\frac{F(A/2,A/2)}{T_{0}} =& -\frac{F(A/2,A/2)}{T(A)} \cdot (1-kA) \\
=& ln[\frac{Y(A/2,A/2)A^{\tau}}{Y_{0}}] \cdot (1-kA)  \\
=&\frac{\widetilde{a_{v}}}{T_{0}}A- \frac{a_{s}}{T_{0}}A^{2/3}-\frac{a_{c}}{T_{0}}\frac{A(A-2)}{4A^{1/3}}\\
&- \frac{a_{p}}{T_{0}}\frac{\delta}{A^{1/2}}+ A(1-kA)ln(\frac{1}{2}),
\end{split}
\label{eq:eq5}
\end{equation}
where $\widetilde{a_{v}}=a_{v}+\frac{1}{2}(\mu_{n}+\mu_{p})$.
The value of $ln[\frac{Y(A/2,A/2)A^{\tau}}{Y_{0}}]$ on the right of the second equation can be calculated from the isotope yields when the $\tau$ value is fixed.
Therefore none zero values of this equation show the deviation of the mass distribution of $N=Z$ isotopes from the power law distribution of the critical exponent~\cite{Bonasera2008}.
In order to eliminate $Y_{0}$, all isotope yields are normalized by the yield of $^{12}C$~\cite{Bonasera2008,Huang10_1,Huang10_2,Chen10,Huang10_3}. For the first round ($k=0$), the renormalized values of $-\frac{F(A/2,A/2)}{T_{0}}$ from the reconstructed isotope yields are plotted as a function of the isotope mass $A$ using solid squares in Fig.~\ref{fig:fig2}(a). The values of $\widetilde{a_{v}}/T_{0}$, $a_{s}/T_{0}$ and $a_{p}/T_{0}$ are used as free parameters to fit the given $-\frac{F(A/2,A/2)}{T_{0}}$ values, employing Eq.\eqref{eq:eq5}. A typical search result is shown by open squares in Fig.~\ref{fig:fig2}(a) for the case of the reconstructed isotope yields at the first round $(k=0)$. Similar quality results are obtained for the AMD simulated events with the three different interactions.
One should note that the value of $a_p/T_{0}$ makes a small contribution and the contribution is evident as a staggering in the $-F(A/2,A/2)/T_{0}$ versus $A$ plot. Therefore the essential free parameters in this step are $\widetilde{a_{v}}/T$ and $a_{s}/T_{0}$.
The extracted parameter values from both experimental data and AMD simulated events are summarized in Table~\ref{table:parameters} for the first round $(k=0)$ and the final round $(k=0.0022)$.

\begin{table}[ht]
\caption{$a/T_{0}$ and $\Delta \mu/T_{0}$ for the first round (k=0) and the final round (0.0022).} 
\centering 
\begin{tabular}{c c c c c c} 
\hline\hline 
      & $\widetilde{a_{v}}/T_{0}$ & $a_{s}/T_{0}$ & $a_{c}/T_{0}$ & $a_{p}/T_{0}$  & $\Delta \mu /T_{0} ^{~a}$ \\ [.5ex] 
\hline 
k=0.0 & & & & &  \\
g0    &    1.15       &    0.0       & $1.82\times 10^{-1}$ & $5.58\times 10^{-1}$ & $6.04\times 10^{-1}$\\
g0AS  &    1.10       &    0.0       & $1.64\times 10^{-1}$ & $6.91\times 10^{-1}$ & $4.80\times 10^{-1}$\\
g0ASS &    1.07       &    0.0       & $1.45\times 10^{-1}$ & $8.98\times 10^{-1}$ & $4.32\times 10^{-1}$\\
Exp.  &    1.08       &    0.0       & $1.44\times 10^{-1}$ & $1.13\times 10^{-1}$ & $6.26\times 10^{-1}$\\
\hline 
k=0.0022 & & & & &  \\
g0    &    1.09       &    0.0       & $1.67\times 10^{-1}$ & $6.34\times 10^{-1}$ & $6.26\times 10^{-1}$\\
g0AS  &    1.04       &    0.0       & $1.50\times 10^{-1}$ & $7.40\times 10^{-1}$ & $5.05\times 10^{-1}$\\
g0ASS &    1.01       &    0.0       & $1.32\times 10^{-1}$ & $9.68\times 10^{-1}$ & $4.51\times 10^{-1}$\\
Exp.  &    1.01       &    0.0       & $1.26\times 10^{-1}$ & $1.72\times 10^{-1}$ & $6.76\times 10^{-1}$\\
\hline 
\end{tabular}
\footnote {$\Delta \mu /T_{0}$ values are taken from the step (1).}
\label{table:parameters} 
\end{table}

In the step (3) Eq.\eqref{eq:eq_MFM} is applied to yields of all isotopes with $N=Z$ and $N \ne Z$. From Eq.\eqref{eq:eq_MFM} $a_{sym}/T_{0}$ and $\Delta \mu / T_{0} = (\mu_n - \mu_p)/T_{0} $ values can be related to the modified free energy, $\frac{\Delta F(N,Z)}{T_{0}}$ as
\begin{equation}
\begin{split}
\frac{\Delta F(N,Z)}{T_{0}}=& \frac{a_{sym}}{T_{0}}\frac{(N-Z)^{2}}{A}-\frac{\Delta \mu}{2T_{0}}(N-Z),\\
\end{split}
\label{eq:eq7}
\end{equation}
where $\frac{\Delta F(N,Z)}{T_{0}}$ is the free energy relative to the temperature, $\frac{F(N,Z)}{T_{0}}$, subtracted by the calculated contributions of the volume, surface, Coulomb and paring terms, using the parameters in Table~\ref{table:parameters}. Resultant $\frac{\Delta F(N,Z)}{T_{0}}$ values are shown by symbols in Fig.~\ref{fig:fig2}(b). They exhibit quadratic shapes with the minimum values close to zero, indicating the $N/Z$ of the fragmenting source is close to 1.
The fluctuation around zero for $N=Z$ isotopes reflects the deviations between the data and the fit points in Fig.~\ref{fig:fig2}(a).

In this step, the $a_{sym}/T_{0}$ and the $\Delta \mu / T_{0}$ values are optimized. Since the $\Delta \mu / T_{0}$ values are extracted from the step(1), the optimization is made for each isotope around the values in the fifth column of Table~\ref{table:parameters} within a small margin. The $a_{sym}/T_{0}$ values are extracted from the quadratic curvature of the isotope distribution for each given $Z$ and plotted in Fig.~\ref{fig:fig2}(c) separately for the AMD simulated events with the g0, g0AS and g0ASS interactions, together with those from the reconstructed isotope yields.

For the first round ($k=0$), the extracted $a_{sym}/T_{0}$ values roughly parallel each other and show a slight increase as $Z$ increases in average for all cases, even though they fluctuate around the average trend, especially for those from the experimentally reconstructed yields.

\section*{III. Characteristic properties of the fragmenting source}

\begin{figure}
\includegraphics[scale=0.45]{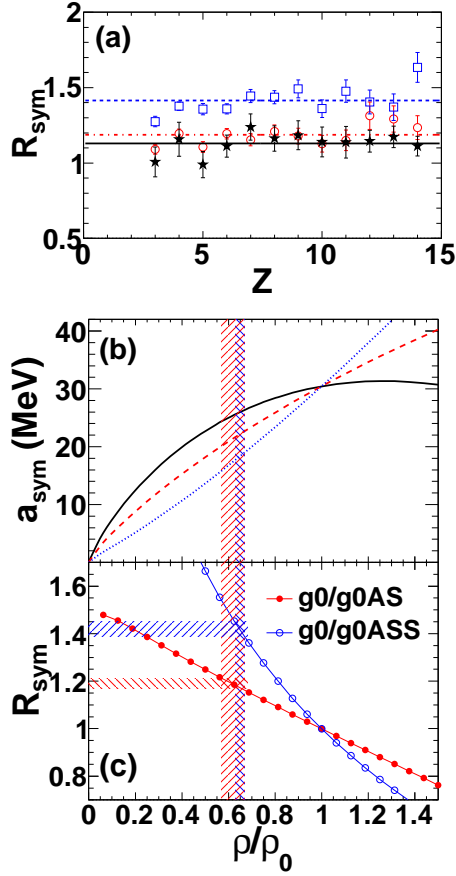}
\caption{\footnotesize (color online)
(a) The ratios of the $a_{sym}/T_{0}$ values shown in Fig.~\ref{fig:fig2}(c), circles for g0/g0AS and squares for g0/g0ASS and stars for g0/Rec..
(b) Symmetry energy coefficient versus density used in the  simulations. Solid curve(g0), dashed (g0AS) and dotted (g0ASS) (c) The ratio of the symmetry energy coefficient in (b). The shaded horizontal lines are the ratios extracted in (a) and the vertical shaded area is the density region corresponding the ratios. Two different shadings are used for the two ratio values. All values are evaluated for the first round $(k=0)$. The ratio and density values are given in Table~\ref{table:data_results}.
}
\label{fig:fig3}
\end{figure}

In order to determine the density and temperature at the time of the fragment formation, the parallel behavior of the observed $a_{sym}/T_{0}$ values in Fig.~\ref{fig:fig2}(c) is utilized.
As suggested in Ref.~\cite{Ono03}, the observed differences are attributed to the difference of the symmetry energy at the density at the time of the fragment formation. The ratios of the $a_{sym}/T_{0}$ values between g0, g0AS, g0ASS and the experimental values for the first round are shown in Fig.~\ref{fig:fig3}(a). The ratios show flat distributions as a function of $Z$ for all cases. The extracted average ratio values are shown by lines in the figure for each ratio and the values are given in the first column of Table~\ref{table:data_results}. In Fig.~\ref{fig:fig3}(b) the symmetry energy coefficient is plotted as a function of the density for the three interactions used in the calculations and in Fig.~\ref{fig:fig3}(c) their ratios, $R_{sym} = a_{sym}(g0)/a_{sym}(g0AS)$ and $R_{sym} = a_{sym}(g0)/a_{sym}(g0ASS)$, are plotted. Using the ratio values determined from Fig.~\ref{fig:fig3}(a) and the density dependence of the $R_{sym}$ values in Fig.~\ref{fig:fig3}(c), the implied densities of the fragmenting sources are indicated by the shaded vertical areas shown in Fig.~\ref{fig:fig3}(c). The extracted density values for each case are given in the second column of Table~\ref{table:data_results}. Assuming that the nucleon density should be same for the three different interactions used, the nucleon density of the fragmenting source is determined from the overlap of the extracted values. This assumption is reasonable for the violent collisions because the nucleon density is mainly determined by the stiffness of the EOS and not by the density dependence of the symmetry energy term. From the overlapped density area in Figs.~\ref{fig:fig3}(c), $\rho/\rho_0 = 0.65 \pm 0.02$ is extracted as the density at the time of the fragment formation.
This overlapped density value is also assigned to the experimental density~\cite{Lin14_1,Lin14_2}. The corresponding symmetry energy values at that density are extracted for the three different interactions from Fig.~\ref{fig:fig3}(b).
The experimental symmetry energy, $a_{sym}(Exp)$ is calculated from the average value of $R_{sym}(Exp)$ shown by the full line in Fig.~\ref{fig:fig3}(a), and $a_{sym}(g0)$ at the obtained density from the AMD events, $\rho/\rho_0=0.65\pm 0.02$, as $a_{sym}(Exp)=a_{sym}(g0)/R_{sym}(Exp)$.
This operation is under the assumption that the system temperatures are almost identical from the AMD events and the experimental reconstructed isotope yields~\cite{Lin14_1,Lin14_2}. Their $a_{sym}$ values are given in the third column of Table~\ref{table:data_results}.

\begin{table}[ht]
\caption{symmetry energy and $\rho/\rho_{0}$ from the first round (k=0)and the final round (0.0022).} 
\centering 
\begin{tabular}{c c c c c} 
\hline\hline 
k&int& $R_{sym}$ & $\rho/\rho_{0}$ & $a_{sym}$ (MeV)\\ [0.5ex] 
\hline 
k=0.0 & & & & \\
&g0    &               &               & 26.0$\pm$0.4 \\
&g0/g0AS       & 1.19$\pm$0.02 & 0.62$\pm$0.05 &              \\
&g0AS  &               &               & 21.4$\pm$1.3 \\
&g0/g0ASS      & 1.42$\pm$0.03 & 0.65$\pm$0.02 &              \\
&g0ASS &               &               & 19.4$\pm$0.7 \\
&g0/Exp.       & 1.13$\pm$0.02 & 0.65$\pm$0.02 &              \\
&Exp.  &               &               & 23.0$\pm$0.6 \\
\hline 
k=0.0022 & & & &\\
&g0    &               &               & 26.0$\pm$0.4 \\
&g0/g0AS      & 1.19$\pm$0.02 & 0.62$\pm$0.05 &              \\
&g0AS  &               &               & 21.4$\pm$1.3 \\
&g0/g0ASS      & 1.42$\pm$0.03 & 0.65$\pm$0.02 &              \\
&g0ASS &               &               & 18.4$\pm$0.7 \\
&g0/Exp.       & 1.13$\pm$0.02 & 0.65$\pm$0.02 &              \\
&Exp.  &               &               & 23.1$\pm$0.6 \\
\hline 
\end{tabular}
\label{table:data_results} 
\end{table}

\begin{figure}
\includegraphics[scale=0.45]{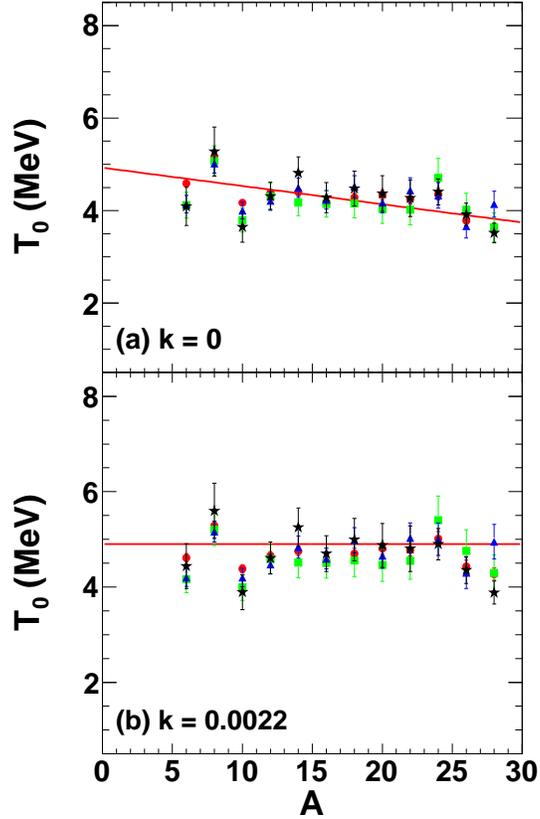}
\caption{\footnotesize (color online)
(a) $T_{0}$ as a function of the isotope mass $A$ for the first round ($k=0$). Solid symbols are same as those in Fig.\ref{fig:fig2}(c). The line is the linear fit of the AMD results. (b) $T_{0}$ as a function of the isotope mass $A$ for the final round (k=0.0022). Same symbol notations are used as (a).}
\label{fig:fig4}
\end{figure}

\begin{figure}
\includegraphics[scale=0.45]{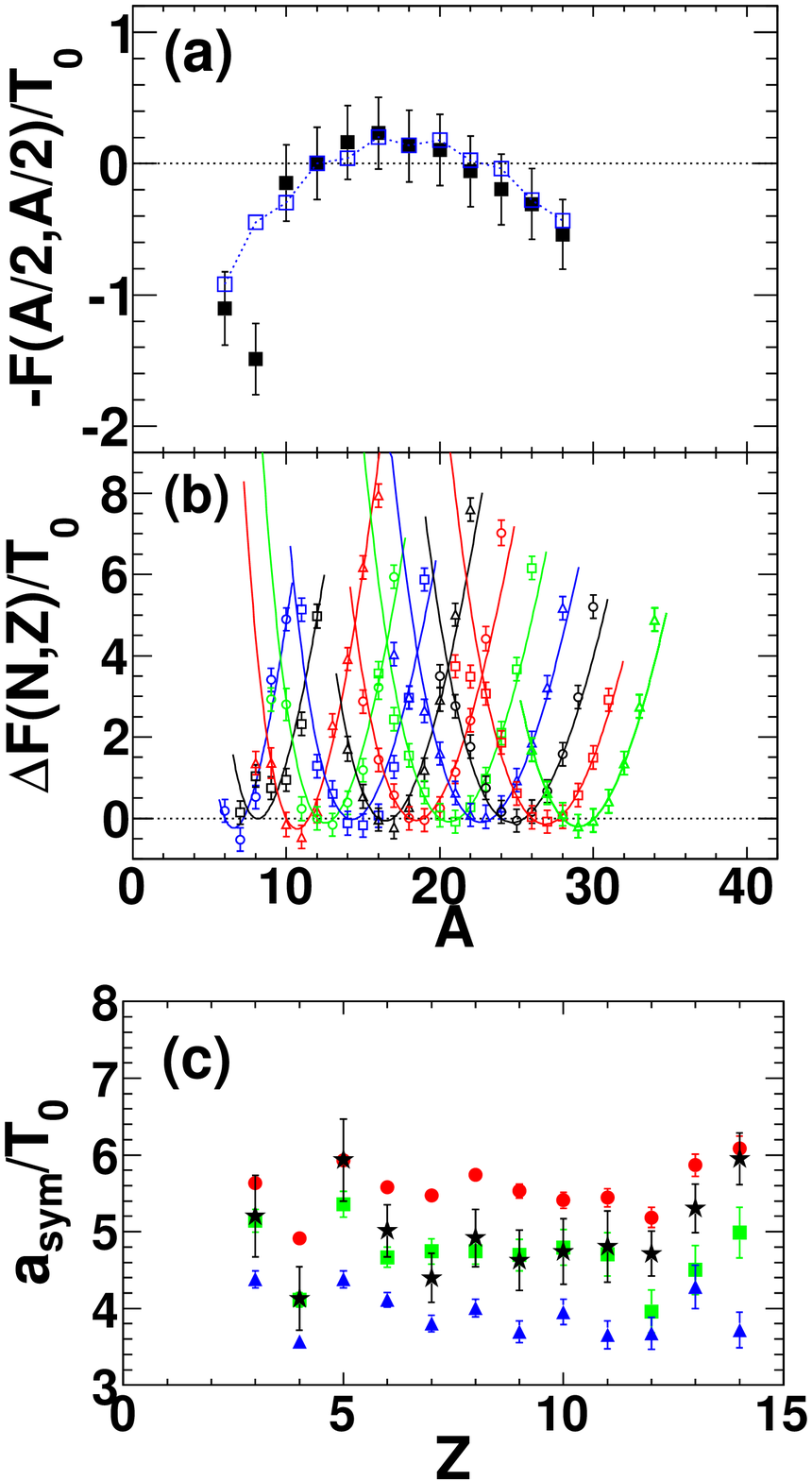}
\caption{\footnotesize (color online)
Same as Fig.\ref{fig:fig2}, but for k=0.0022.
}
\label{fig:fig5}
\end{figure}

Once the symmetry energy value is determined for the individual cases, the temperature $T_0$ can be calculated as $T_{0} = a_{sym}/(a_{sym}/T_{0})$. The extracted $T_0$ values from the reconstructed isotope yields and the AMD events are shown as a function of $A$ by different solid symbols for the first round ($k=0$) in Fig.~\ref{fig:fig4}(a), under the assumption of $A \sim 2Z$. Temperature values extracted from the experimentally reconstructed yields and the AMD simulated events with three different interactions agree with each other within the error bars. The larger errors in these plots, comparing to those in Fig.~\ref{fig:fig2}(c), originate from the errors of $a_{sym}$ and $a_{sym}/T_{0}$ which are shown in the third column of Table~\ref{table:data_results} and Fig.~\ref{fig:fig1}(c), respectively. The extracted temperature values show a monotonic decreasing trend as $A$ increases from $\sim 5 $ MeV at $A =6$ to $\sim 3.5 $ MeV at $A = 28$. From the linear fit, $T_0 =4.9( 1 - 0.008A)$, is determined for the first round.

The iteration is repeated three times in this work. The same plots as Fig.\ref{fig:fig2}, but with the $k$ value for the final (third) round, $k=0.0022$, are shown in Fig.\ref{fig:fig5} and the extracted parameters are also given in Table~\ref{table:parameters}. A very similar quality of results is obtained between those of the first round $(k=0)$ and of the final round $(k=0.0022)$.
The extracted $T_{0}$ values are shown in Fig.~\ref{fig:fig4}(b) for the final round ($k=0.0022$). All extracted $T_0$ values show a flat distribution and therefore the iteration stops at this round. The extracted $T_{0}$ values from the reconstructed isotope yields and the AMD events agree with each other within the error bars and $T_{0}=5.0 \pm 0.4 $ MeV is extracted, where the error is calculated from the standard deviation.

The extracted density and symmetry energy in the different iteration round are very similar as seen in Table~\ref{table:data_results}, even though the parameter values in Table~\ref{table:parameters} are 5 to 10\% different in some cases. This indicates that the extracted density, symmetry energy and temperature values in Table~\ref{table:data_results} are quite stable in the iteration procedures.
All parameters extracted in this work are also consistent to those in the previous works~\cite{Lin14_1,Lin14_2}, in which a simpler method is employed to evaluate $a_{sym}/T_{0}$ values.

\section*{IV. Discussion}

\begin{figure}
\includegraphics[scale=0.45]{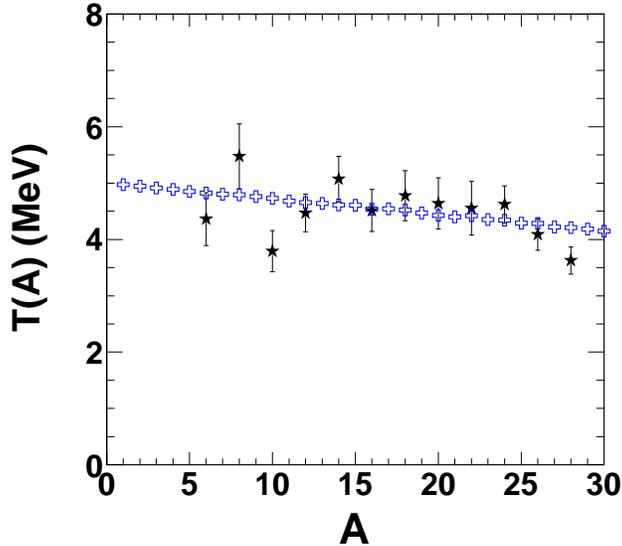}
\caption{\footnotesize (color online)
Mass dependent temperature T(A) as a function of the isotope mass A. Starts are calculated from $T_{0}$ in Fig.\ref{fig:fig4}(b) times $(1-kA)$ for the final round $(k=0.0022)$. Crosses are the results of the Monte Carlo simulation of the thermal motion under the momentum conservation with $T_{0}=5.0$ MeV.
}
\label{fig:fig6}
\end{figure}

In order to study the observed slope in the apparent temperature, a simple Monte Carlo model is employed, following Ref.~\cite{Liu14}.
Under a thermal equilibrium condition, the thermal motion with velocity $v^{th}_{i}$, where $i=x,y,z$, is expressed by a Maxwell-Boltzmann distribution as
\begin{equation}
\begin{split}
v^{th}_{i}\sim exp[-\frac{(v_{i}^{th})^{2}}{2 \cdot (T_0/A)}],
\end{split}
\label{eq:eq_th}
\end{equation}
where $T_0$ is the input parameter in the model. Fragments are generated by a percolation model for a system with mass 180 ($6 \times 6 \times 5$)~\cite{Staufer1979}. $T_{0}=5.0$ MeV is used, which represents the thermal temperature of nucleons in the model. More than a hundred million events are generated. In order to require the momentum conservation in the fragmenting system, the events which satisfy the condition of $|\sum_j{m_j\overrightarrow{v(j)}| \le 100}$ MeV/c are selected as an approximation of the momentum conservation, $\sum_j{m_j\overrightarrow{v(j)}=0}$.
The temperature value from this Monte Carlo simulation is evaluated utilizing a fluctuation thermometer under a classical momentum distribution. Detailed descriptions about this classical fluctuation thermometer can be found in Ref.~\cite{Wuenschel2010}.
The results are shown by open crosses in Fig.\ref{fig:fig6}.
The slight mass dependence of the temperature as $A$ increases is observed, which originates from the requirement of the momentum conservation. When the thermal motion is distributed equally to the fragments in a finite system according to Eq.\eqref{eq:eq_th}, the larger fragments result in larger momentum and their momentum fluctuation becomes larger. Therefore the larger fragments become less probable to satisfy the requirement of the momentum conservation for an equal distribution of the thermal motion among the fragments.
The mass dependent temperature $T(A)$ from the experimentally reconstructed yields is plotted in Fig.\ref{fig:fig6} together with the results of this simple Monte Carlo simulation. The experimental trend is rather well reproduced.
As a conclusion, the observed mass dependence of the temperature is well explained by an equal distribution of the thermal motion of $T=5.0$ MeV under the momentum conservation, which is closely related to the size of the system. In fact, in Ref.~\cite{Liu14} the same procedures above are applied to the $^{40}Ca+^{40}Ca$ reaction and $k=0.007$ is obtained, which is $\sim3$ times larger than that of the present case. That is because when the system becomes larger, the fragments suffer less restriction under the momentum conservation, comparing to those in the smaller system.

\section*{V. SUMMARY}

An improve method to extract the symmetry energy coefficients relative to the temperature, $a_{sym}/T_{0}$, and a self-consistent determination of the density, temperature and symmetry energy of the fragmenting system are presented. Using the improved method based on the MFM model, $a_{sym}/T_{0}$ values are extracted, utilizing all of the reconstructed hot isotope yields and the AMD simulated events with the Gogny interaction with three different density dependencies of the symmetry energy term. The extracted $a_{sym}/T_{0}$ values show a monotonic increase trend as isotope mass $A$ increases. The AMD results show that they are more or less in parallel each other. This parallel behavior is interpreted as the reflection of the different symmetry energy values at a given density and temperature at the time of the fragmentation of the system. Using this correlation, the density value is first determined as $\rho/\rho_{0} = 0.65 \pm 0.2 $ for the fragmenting system in the experiment. Utilizing this density, the symmetry energies are evaluated in a self-consistent manner for each AMD simulation. The extracted symmetry energy value for the experimentally reconstructed isotope yields is $a_{sym}= 23.1 \pm 0.6$ MeV. Using the extracted symmetry energy values, the temperature values are calculated for the reconstructed isotope yields and those of the AMD simulated events. They agree each other within the error bars and show a slight linear decrease as $A$ of the fragments increases. For the final (third) round of the iteration, $T(A)=T_{0}(1-kA)$ is obtained, where $=5\pm0.4$ MeV and $k=0.0022$. In the different iteration stages, the extracted density and symmetry energy agree within the error bars, indicating that these extracted values do not depend so much on the optimized parameter values. Using a simple Monte Carlo simulation, the mass dependence of the apparent temperature is well explained by an equal distribution of the thermal motion to different size of fragments under the momentum conservation, indicating that the mass dependence of the apparent temperature originates from the system size effect.

\section*{Acknowledgments}

The authors thank to the operational staff in the cyclotron Institute, Texas A\&M University, for their support during the experiment. The experiment is supported by the U.S. Department of Energy under Grant No. DE-FG03-93ER40773 and the Robert A. Welch Foundation under Grant A0330. This work is also supported by the National Natural Science Foundation of China (Grants No. 11075189, No. 11205209 and No. 11105187) and 100 Persons Project (Grants No. 0910020BR0 and No. Y010110BR0), ADS project 302 (Grants No. Y103010ADS) of the Chinese Academy of Sciences. One of the authors (R.W.) thanks the program of the visiting professorship of senior international scientists of the Chinese Academy of Science" (Grant No. 2012T1JY3-2010T2J22) for their support.









\end{document}